\documentclass[prd,aps,preprint,tightenlines]{revtex4}
\usepackage{epsfig}
\newcommand{\insertfig}[2]{\mbox{\epsfxsize=#1cm \epsfbox{#2.eps}}}
\begin{document}
\title{Quark Imaging in the Proton \\ Via Quantum
Phase-Space Distributions}
\author{A.V. Belitsky}
\email{belitsky@physics.umd.edu}
\affiliation{Department of Physics, University of Maryland \\
College Park, Maryland 20742 \\
\\
Institute for Nuclear Theory, University of Washington \\
Box 351550, Seattle, WA, 98195}
\author{Xiangdong Ji}
\email{xji@physics.umd.edu}
\affiliation{Department of Physics, University of Maryland \\
College Park, Maryland 20742 \\
\\
Institute for Nuclear Theory, University of Washington \\
Box 351550, Seattle, WA, 98195}
\author{Feng Yuan}
\email{fyuan@physics.umd.edu}
\affiliation{Department of Physics, University of Maryland \\
College Park, Maryland 20742 \\
\\
Institute for Nuclear Theory, University of Washington \\
Box 351550, Seattle, WA, 98195}

\date{\today}

\begin{abstract}

We develop the concept of quantum phase-space (Wigner)
distributions for quarks and gluons in the proton. To appreciate
their physical content, we analyze the contraints from special
relativity on the interpretation of elastic form factors, and
examine the physics of the Feynman parton distributions in the
proton's rest frame. We relate the quark Wigner functions to the
transverse-momentum dependent parton distributions and
generalized parton distributions, emphasizing the physical role
of the skewness parameter. We show that the Wigner functions
allow to visualize quantum quarks and gluons using the language
of the classical phase space. We present two examples of the
quark Wigner distributions and point out some model-independent
features.

\end{abstract}
\maketitle

\section{Introduction}

In exploring the microscopic structure of matter, there are two
frequently-used approaches. First, the spatial distribution of
matter (or charge) in a system can be probed through elastic
scattering of electrons, or photons, or neutrons, etc. The
physical quantity that one measures is the elastic form
(structure) factors which depend on three-momentum transfer to
the system. The Fourier transformation of the form factors
provides direct information on the spatial distributions. The
well-known examples include the study of charge distribution in
an atom and the atomic structure of a crystal. The second approach
is designed to measure the population of the constituents as a
function of momentum, or the momentum distribution, through
knock-out scattering. Here the well-known examples include the
nucleon distributions in nuclei measured through quasi-elastic
electron scattering, and the distribution of atoms in a quantum
liquid probed through neutron scattering. The scattering cross
section sometimes depends on the reaction dynamics which must be
understood before the momentum distribution can be extracted.

Both approaches are complementary, but bear similar drawbacks. The
form factor measurements do not yield any information about the
underlying dynamics of the system such as the speed of the
constituents, whereas the momentum distribution does not give any
information on the spatial location of the constituents. More
complete information about the microscopic structure lies in the
correlation between the momentum and coordinate spaces, i.e., to
know where a particle is located and, at the same time, with what
velocity it travels. This information is certainly attainable for
a classical system for which one can define and study the
phase-space distribution of the constituents. For a quantum
mechanical particle, however, the notion of a phase-space
distribution seems less useful because of the uncertainty
principle. Nonetheless, the first phase-space distribution in
quantum mechanics was introduced by Wigner in 1932 \cite{Wig32},
and many similar distributions have been studied
thereafter. These distributions have been used for various
purposes in very diverse areas such heavy-ion collisions, quantum
molecular dynamics, signal analysis, quantum information, optics,
image processing, non-linear dynamics, etc.\cite{HilOcoScuWig84}
In certain cases, the Wigner distributions can even be measured
directly in experiments
\cite{VogRis89,SmiBacRayFar93,BanRadWodKra99}, providing much
information about the dynamics of a system.

The main interest of this paper is about the internal structure
of the proton (or neutron), for which the underlying fundamental
theory is quantum chromodynamics (QCD). With some changes to
accommodate the relativistic nature of the problem, both
experimental approaches alluded to above have been successfully
used to unravel its quark and gluon structure: The elastic form
factors of the proton have been measured since the 1950s and, at
low-momentum transfer ($\le$ nucleon mass $M_N$) where the
nucleon recoil effects are small, the three-dimensional (3D)
Fourier transformation of these form factors can be interpreted
as spatial charge and current distributions of quarks
\cite{Sac60}. Feynman parton distributions, measurable in
high-energy inelastic scattering such as deep-inelastic
scattering (DIS) and Drell-Yan process, have a simple
interpretation as the momentum distributions of the quarks and
gluons in the infinite momentum frame (IMF) \cite{Fey72}.
However, the notion of correlated position and momentum
distributions of quarks and gluons has not been systematically
investigated in the field, although it is clear that the physics
of a phase-space distribution must be very rich.

In this paper, we explore to what extent one can construct
physically-interesting and experimentally-measurable phase-space
distributions in QCD, and what information it contains about the
QCD parton dynamics. [A brief account of some of the results can
be found in Ref. \cite{Ji03}, see also \cite{Bel03}.] To facilitate
the construction, we examine the uncertainty in the traditional
interpretation of electromagnetic form factors due to relativity,
and analyze the physical content of the Feynman parton distributions
in the rest frame of the proton. We then introduce the phase-space
Wigner distributions for the quarks and gluons in the proton, which
contain most general one-body information of partons,
corresponding to the full one-body density matrix in technical
terms. After integrating over the spatial coordinates, one
recovers the familiar transverse-momentum dependent parton
distributions \cite{Col03}. On the other hand, some reduced
version of the distributions is related, through a specific
Fourier transformation, to the generalized parton distributions
(GPDs) which have been studied extensively in the literature in
recent years
\cite{Mue94,Ji97,Rad97,Ji98,GoePolVan01,Rad01,Bel01}. Roughly
speaking, a GPD is a one-body matrix element which combines the
kinematics of both elastic form factors and Feynman parton
distributions, and is measurable in hard exclusive processes.
Therefore, the notion of phase-space distribution provides a new
3D interpretation of the GPDs in the rest frame of the proton.
There are other interpretations of the GPD in the literature
which are made in IMF and impact parameter space
\cite{Sop77,Bur00,RalPir02,BelMul02,Die02}.

 The presentation of the paper is as follows. In Section II, we
examine the constraints on the physical interpretation of the
form factors and parton distributions from relativistic effects,
anticipating their extension to a full phase-space distribution.
In Section III, we first briefly summarize the main features of a
quantum mechanical Wigner distribution, then introduce the
quantum phase-space distributions for the quarks and gluons in a
rest-frame proton. In Section IV, we exhibit the spatial 3D
images of quarks generated from slicing the quantum phase-space
(Wigner) distributions at different Feynman momentum and comment
on their general features. Section V contains the summary and
conclusion.

\section{Relativity Constraint On Interpretation of Form Factors
and Parton Distributions}

In the literature, the quantum phase-space distributions have
been mostly applied to non-relativistic systems. For the proton,
however, relativity is essential. In measuring the elastic form
factors of the proton, the momentum transfer to the system can
easily exceed the rest mass, resulting a large recoil and Lorentz
contraction. The quarks and gluons inside the proton follow
relativistic dynamics. Moreover, when a quark is struck in a DIS
experiment, it travels along the light-cone: the trajectory of an
extreme-relativistic particle. Therefore, to develop a
phase-space distribution of the proton, we must examine to what
extent the notion actually makes sense for relativistic systems.

In the first subsection, we examine the textbook interpretation
of the electromagnetic form factors of the proton, reminding the
reader that there are intrinsic ambiguities in the interpretation.
We emphasize, however, that different ways of the interpreting the
form factors can be regarded as different choices of schemes. When
used consistently, one scheme is in principle as good as any
other. The degree of scheme dependence depends on $1/(MR)$, where
$M$ is the mass and $R$ is some kind of radius, which is $~$1/4
for the proton.

In the second subsection, we consider the Feynman parton
distributions, most-commonly interpreted as the momentum densities
in IMF. Since the notion of a phase-space distribution is meant
for a proton in its {\it rest frame}, and since the distribution
should be reduced to the Feynman distribution after integrating
out the spatial coordinates, we are compelled to examine the
physics of latter in the static system of coordinates. In
particular, we need to understand the meaning of Feynman momentum
$x$ in that context. This can be achieved by introducing the
so-called {\it spectral function}---the correlated momentum and
energy distribution of the constituents---familiar in
non-relativistic many-body physics. In the process, we find that
the separation between particle and antiparticle, familiar in the
IMF, disappears: One can only keep track of the creation and
annihilation of fermion quantum numbers, such as the electric
charge.

\subsection{The Proton Form Factors and Scheme-Dependent Charge Distributions}

The electromagnetic form factors are among the first measured and
mostly studied observables of the nucleon. They are defined as the
matrix elements of the electromagnetic current between the nucleon
states of different four-momenta. Because the nucleon is a spin
one-half particle, the matrix element defines two form factors,
\begin{equation}
\langle p_2 | j_\mu (0) |p_1 \rangle = \bar U (p_2) \left\{ F_1(
q^2) \gamma_\mu + F_2 (q^2) \frac{i \sigma_{\mu\nu} q_\nu}{2M_N}
\right\} U (p_1) \, ,
\end{equation}
where $F_1$ and $F_2$ are the well-known Dirac and Pauli form
factors, respectively, depending on the momentum transfer $q = p_2
- p_1$, and $U(p)$ is nucleon spinor normalized as
$\overline{U}(p) U (p) = 2 M_N$.

Since the beginning, it has been known that the physical
interpretation of the nucleon form factors is complicated by
relativistic effects \cite{YenRavLev57}. Consider a system of
size $R$ and mass $M$. In relativistic quantum theory, the system
cannot be localized to a precision better than its Compton
wavelength $1/M$. Any attempt to do this with an external
potential will result in creation of particle-antiparticle pairs.
As a consequence, the static size of the system cannot be defined
to a precision better than $1/M$. If $R \gg 1/M$, which is the
case for all non-relativistic systems, the above is not a
significant constraint. One can probe the internal structure of
the system with a wavelength ($1/|\vec{q}|$) comparable to or even
much smaller than $R$, but still large enough compared to $1/M$ so
that the probe does not induce an appreciable recoil. A familiar
example is the hydrogen atom for which $R M_H \sim M_H/(m_e
\alpha_{\rm em}) \sim 10^5$, and the form factor can be measured
through electron scattering with momentum transfer $|\vec{q}| \ll
M_H$.

When the probing wavelength is comparable to $1/M$, the form
factors are no longer determined by the internal structure alone.
They contain also the dynamical effects of Lorentz boosts because
the initial and final protons have different momenta. In
relativistic quantum theory, the boost operators involve
nontrivial dynamical effects which result in the nucleon wave
function being different in different frame (in the usual instant
form of quantization). Therefore in the region $|\vec{q}| \sim M$,
the physical interpretation of the form factors is complicated
because of the entanglement of the internal and the
center-of-mass motions in relativistic dynamics. In the limit
$|\vec{q}| \gg M$, the former factors depend almost entirely on
the physical mechanism producing the overall change of the proton
momentum. The structural effect involved is a very small part of
the nucleon wave function (usually the minimal Fock component
only).

For the nucleon, $M_N R_N \sim 4$. Although much less certain
than in the case of the hydrogen atom, it seems still sensible to
have a rest-frame picture in terms of the electromagnetic form
factors, so long as one keeps in mind that equally justified
definitions of the nucleon sizes can differ by $\sim 1/M_N (R_N
M_N)$. For example, the traditional definition of the proton
charge radius in terms of the slope of the Sachs form factor $G_E
(q^2)$ is 0.86 fm. On the other hand, if one uses the slope of
the Dirac form factor $F_1$ to define the charge radius, one gets
0.79 fm, about 10\% smaller. The form factors at $|\vec{q}| \ge
M_N \sim 1$ GeV cannot be interpreted as information about the
internal structure alone.

To further clarify the uncertainty involved in the interpretation
of the electromagnetic form factors, let us review the textbook
explanation offered originally by Sachs \cite{Sac60}.

To establish the notion of a static (charge) distribution, one
needs to create a wave-packet representing a proton localized at
$\vec{R}$
\begin{equation}
| \vec{R} \rangle = \int \frac{d^3\vec{p}}{(2\pi)^3} \, {\rm
e}^{i \vec{\scriptstyle p} \cdot \vec{\scriptstyle R}} \,
{\mit\Psi} (\vec{p}) | \vec{p} \rangle \, ,
\end{equation}
where the plane-wave state $|\vec{p}\rangle$ is normalized in a
relativistic-covariant manner $\langle \vec{p}_2 | \vec{p}_1
\rangle = 2 E_{\vec{\scriptstyle p}_1} (2\pi)^3 \delta^{(3)}
(\vec{p}_1 - \vec{p}_2)$, and ${\mit\Psi} (\vec{p})$ is the
momentum space profile normalized as $2 \int d^3\vec{p} \,
E_{\vec{\scriptstyle p}} |{\mit\Psi} (\vec{p})|^2 = (2\pi)^3$.
The wave-packet is not an eigenstate of the free Hamiltonian.
Therefore, as time progresses, the wave packet will spread. The
characteristic spreading time is proportional to $\langle
M_N/\vec{p}^2 \rangle$ which is long for a non-relativistic
system. But for a relativistic particle, the spread could happen
much faster compared to the characteristic time-scale of a
weakly-interacting probe.

Having localized the wave-packet at $\vec{R} = 0$, we can
calculate, for example, the charge distribution in the wave-packet
\begin{equation}
\rho (\vec{r}) = \langle \vec{R} = 0| j_0 (\vec{r}) | \vec{R} = 0
\rangle \, ,
\end{equation}
where $\vec{r}$ measures the relative distance to the center
$\vec{R} = 0$. Taking its Fourier transform, one gets
\begin{eqnarray}
F (\vec{q}) &\equiv&  \int d^3 \vec{r} \, {\rm e}^{i
\vec{\scriptstyle q} \cdot \vec{\scriptstyle r}} \rho (\vec{r})
\nonumber \\
&=& \int \frac{d^3 \vec{p}}{(2\pi)^3} {\mit\Psi}^* \left( \vec{p}
+  \vec{q}/2 \right) {\mit\Psi} \left( \vec{p} - \vec{q}/2
\right) \left\langle \vec{p} + \vec{q}/2 \right| j_0 (0) \left|
\vec{p} - \vec{q}/2 \right\rangle \, ,
\end{eqnarray}
where we have changed the momentum integration variables, with
$\vec{p}$ representing the average momentum of the initial and
final protons. It is important to point out that the resolution
momentum $\vec{q}$ is now linked to the difference in the initial
and final state momenta. In non-relativistic quantum systems,
because of the large masses, the momentum transfer causes little
change in velocity, and hence the initial and final states have
practically the same intrinsic wave functions. In relativistic
systems, this is the origin of the difficulty in interpreting the
form-factor: we do not have a matrix element involving the same
intrinsic proton state.

To remove the effects of the wave packet, the necessary condition
on ${\mit\Psi} (\vec{p})$ is that the coordinate-space size of the
wave-packet must be much smaller than the system size $|\delta
\vec{r}| \ll R_N$. Furthermore, the probing wave length, or
resolution scale, must also be large compared with the size of the
wave-packet $\delta \vec{r}\sim 1/\vec{p} \ll 1/\vec{q}$. Then
one can ignore $\vec{q}$-dependence in ${\mit\Psi}$, so that
${\mit\Psi} \left( \vec{p} \pm \frac{1}{2} \vec{q} \right)
\approx {\mit\Psi} (\vec{p})$
\begin{equation}
F (\vec{q}) = \int \frac{d^3 \vec{p}}{(2\pi)^3} | {\mit\Psi}
(\vec{p}) |^2 \left\langle \vec{p} + \vec{q}/2 \right| j_0 (0)
\left| \vec{p} - \vec{q}/2 \right\rangle \, .
\end{equation}
On the other hand, to be insensitive to the anti-particle degrees
of freedom, the size of the wave packet must be larger than the
proton Compton wave length $|\delta \vec{r}| \gg 1/M_N$. In the
momentum space this corresponds to restrictions on momenta
allowed in the wave packet $|\vec{p}| \ll M_N$. Therefore the
combined constraint on the wave packet profile is $1/R_N \ll
|\vec{q}| \ll  |\vec{p}| \ll M_N$. The extreme limit of the last
inequality yields a wave packet with a zero-momentum nucleon
\begin{equation}
| {\mit\Psi} (\vec{p}) |^2 = \frac{(2 \pi)^3}{2 M_N} \delta^{(3)}
( \vec{p} ) \, ,
\end{equation}
which gives
\begin{equation}
2 M_N F (\vec{q}) = \left\langle  \vec{q}/2 \right| j_0 (0) \left|
- \vec{q}/2 \right\rangle \, .
\end{equation}
This is the matrix element of the charge density in the Breit
frame, and is  $2 M_N G_E (t) w_2^\ast w_1$ where
\begin{equation}
 G_E (t) = F_1 (t) + \frac{t}{4 M_N^2} F_2 (t)
\end{equation}
is the Sachs electric form factor ($t = - \vec{q}^2$) and the Weyl
spinors involved are normalized conventionally by $w^\ast w = 1$.
Hence, we arrive at the textbook interpretation of $G_E$ as a
Fourier transformation of the proton charge distribution.

Likewise, the Sachs magnetic form factor $G_M (t) = F_1 (t) + F_2
(t)$ is obtained from the Breit frame matrix element of the
electric current
\begin{eqnarray}
\left\langle \vec{q}/2 \right| \vec{j} (0) \left| -  \vec{q}/2
\right\rangle = 2 i [ \vec{s} \times \vec{q}] G_M (t) \, ,
\end{eqnarray}
where the three-vector of spin is $\vec{s} = w^\ast_2 \frac{1}{2}
\vec{\sigma} w_1$.

It must be pointed out that the charge and magnetization
distributions thus defined contains the Lorentz contraction
effects along the photon direction $\vec{q}$ when $\vec{q}^2 \gg
4 M_N^2$, which make the proton look like a pancake. Various
prescriptions exist in the literature which have been proposed to
remove the relativity effects and extract the ``intrinsic"
charge/magnetization distributions from the experimental data
\cite{LicPag70,Ji91,Kel02}. However, it is difficult to accomplish
it in a model-independent way.

Since relativity makes the interpretation of the electromagnetic
form factors non-unique, the best one can do is to choose one
particular interpretation and work consistently. For example, when
extracting the proton charge radius from the Lamb shift
measurements, one shall use the same definition as from the
electric form factor. The most frequently-used definition is that
of Sachs, but other schemes are equally good and the scheme
dependence disappears in the limit $MR\rightarrow \infty$. This
is very much like the renormalization scheme dependence of parton
densities due to radiative corrections at finite strong coupling
constant $\alpha_{\rm s}$: Although the $\overline{\rm MS}$ scheme
is the most popular in the literature, one can use the parton
densities in any other scheme to correlate physical observables.
In this paper, we adopt the Sachs interpretation of the form
factors.

Relativistic corrections and Lorentz contraction effects in the
transverse dimensions are found to disappear in an IMF
\cite{Bur00}. There the proton has an infinitely large effective
mass, and hence for physics in the transverse dimensions, we are
back to the non-relativistic case. In particular, one can
localize the proton in the transverse coordinate space with no
recoil corrections. The Dirac form factor $F_1$ is found to be
related to the charge distribution in transverse plane, with
information along the $z$-axis integrated. The price one pays for
eliminating the relativistic effects is the loss of a spatial
dimension.

\subsection{Parton Distributions As Seen in the Rest Frame of the Proton}

Parton distributions were introduced by Feynman to describe
deep-inelastic scattering \cite{Fey72}.  They have the simplest
interpretation in the IMF as the densities of partons in the
longitudinal momentum $x$. In QCD, the quark distribution is
defined through the following matrix element,
\begin{equation}
 q(x) = \frac{1}{2 p^+}
\int \frac{d\lambda}{2\pi}
{\rm e}^{i\lambda x}
\langle p|\overline{\mit{\Psi}}(0)\gamma^+ \mit{\Psi}(\lambda n)
|p\rangle \ , \label{qcdpd}
\end{equation}
where we have used the standard light-cone notation $p^\pm
=(p^0\pm p^3)/\sqrt{2}$, and $n^\mu$ is a vector along the
direction of $(1,0,0,-1)$ and $n \cdot p = 1$. $\mit{\Psi}$ is a
quark field with an associated gauge link extending from the
position of the quark to infinity along the light cone, and hence
is gauge-invariant in non-singular gauges. The renormalization
scale dependence is implicit. In light-cone quantization
\cite{BroPauPin97}, it is easy to get
\begin{eqnarray}
q (x)|_{x > 0} &=& \frac{1}{2 x} \sum_{\lambda =
\uparrow\downarrow} \int \frac{d^2 \vec{k}_\perp}{(2 \pi)^3}
\frac{ \langle p | b_\lambda^\dagger (k^+, \vec{k}_\perp)
b_\lambda (k^+, \vec{k}_\perp) | p \rangle }{ \langle p | p
\rangle }
\, , \nonumber \\
q (x)|_{x < 0} &=& \frac{-1}{2 x} \sum_{\lambda =
\uparrow\downarrow} \int \frac{d^2 \vec{k}_\perp}{(2 \pi)^3}
\frac{ \langle p | d_\lambda^\dagger (k^+, \vec{k}_\perp)
d_\lambda (k^+, \vec{k}_\perp) | p \rangle }{ \langle p | p
\rangle } \, ,
\end{eqnarray}
where $b^\dagger$ and $d^\dagger$ are creation operators of a
quark and an anti-quark, respectively, with longitudinal momentum
$k^+ \equiv x p^+$ and transverse momentum $\vec{k}_\perp$. The
interpretation as parton densities is then obvious.

To construct the quantum phase-space distributions for the quarks,
we need an interpretation of the Feynman densities in the rest
frame. This is because the IMF involves a Lorentz boost along the
$z$-direction which destroys the rotational symmetry of the 3D
space. However, if one works in the rest frame of the proton, the
two quark fields in Eq. (\ref{qcdpd}) are not at the same time.
If one Fourier-expands one of the fields in terms of quark
creation and annihilation operators, the other must be determined
from Heisenberg equation of motion. The result is that the
bi-linear quark operator takes a very complicated expression in
terms of the creation and annihilation operators in the
equal-time quantization.

The physics of the Feynman quark distribution in the rest frame is
made more clear through the notion of the {\it spectral function}
\begin{equation}
S(k) = \frac{1}{2 p^+} \int d^4 \xi {\rm e}^{i k \cdot \xi}
\langle p | \overline{\mit{\Psi}}(0) \gamma^+ \mit{\Psi}(\xi) | p \rangle \ .
\end{equation}
which is the dispersive part of the single-quark Green's function
in the proton. The physical meaning of $S(k)$ can be seen from its
spectral representation,
\begin{eqnarray}
S(k) &= &
\sum_n (2\pi)^4 \delta^{(4)} (p-k-p_n) \langle p|\overline{\mit{\Psi}}_k|n\rangle
\gamma^+\langle n|{\mit\Psi}(0)|p\rangle/2p^+
\nonumber \\
&\sim & \sum_n (2\pi)^4 \delta^{(4)} (p-k-p_n)
|\langle n|{\mit\Psi}_{k+}|p\rangle|^2
\end{eqnarray}
where ${\mit\Psi}_k$ is a Fourier transformation of
${\mit\Psi}(\xi)$: It is the probability of annihilating a quark
(or creating an antiquark) of {\it four}-momentum $k$
(three-momentum $\vec{k}$ and the off-shell energy $E=k^0$) in
the nucleon, leading to an ``on-shell" state of energy-momentum
$p_n=p-k$. The quark here is off-shell because if $p_n$ and $p$
are both ``on-shell", $k^2\ne m_q^2$ in general. [That the
partons are off-shell are in fact also true in the IMF
calculations.] Of course, in QCD $|n\rangle$ is not in the
Hilbert space, but the spectral function itself is still a
meaningful quantity.

Since the quarks are ultra-relativistic, ${\mit\Psi}_k$ contains
both quark and antiquark Fock operators. One cannot in general
separate quark and anti-quark contributions, unlike in the
non-relativistic systems in which only the particle or
antiparticle contribute. In fact, if one expands the above
expression, one finds pair creations and annihilation terms.
However, this is also true for the charge density discussed in the
previous subsection. Therefore we can speak of $S(k)$ as a
distribution of vector charges and currents, but not a particle
density. In nuclear physics where the non-relativistic dynamics
dominates, the nucleon spectral function in the nucleus is
positive definite and can be regarded as a particle density. The
nuclear spectral function is directly measurable through pick-up
and knock-out experiments, in which $E$ and $\vec{k}$ are called
the missing energy and missing momentum, respectively (see for
example \cite{JiMck90}).

It is now easy to see that in the rest frame of the proton, the
Feynman quark distribution is
\begin{equation}
q(x) = \sqrt{2}\int \frac{d^4k}{(2\pi)^4}
\delta (k^0+k^z-xM_N) S(k) \ .
\end{equation}
The $x$ variable is simply a special combination of the off-shell
energy $k^0$ and momentum $k^z$. The parton distribution is the
spectral function of quarks projected along a special direction in
the four-dimensional energy-momentum space. The quarks with
different $k^0$ and $k^z$ can have the same $x$, and moreover, the
both $x>0$ and $x<0$ distributions contain contributions from
quarks and anti-quarks.

To summarize, in the proton rest frame, the quarks are naturally
off-shell, and hence have a distribution in the four-dimensional
energy-momentum space. The Feynman distribution comes from a
reduction of the full distribution along a special direction.

\section{Quantum Phase-Space (Wigner) Distributions}

In classical physics, a state of a particle is specified by its
position $\vec{r}$ and momentum $\vec{p}$. In a gas of classical
particles, the single-particle properties are described by a
phase-space distribution $f(\vec{r},\vec{p})$ representing the
density of particles at a phase-space point $(\vec{r},\vec{p})$.
Time evolution of the distribution is governed by the Boltzmann
equation, or Liouville equation if the particles are not
interacting.

In quantum mechanics, position and momentum operators do not
commute and hence, in principle, one cannot talk about a joint
momentum and position distribution of particles. Indeed the
quantum mechanical wave functions depend on either spatial
coordinates or momentum, but never both. Nonetheless, Wigner
introduced the first quantum phase-space distribution just a few
years after quantum mechanics was formulated \cite{Wig32}. It is
not positive definite and hence cannot be regarded as a
probability distribution. However, it reduces to the
positive-definite classical phase-space distribution in
$\hbar\rightarrow 0$ limit. The sign oscillation in the
phase-space is necessary to reproduce quantum interference. The
Wigner distribution contains the complete single-particle
information about a quantum system (equivalent to the full
single-particle density matrix), and can be used to calculate any
single-particle observable through classical-type phase-space
averages.

In this section, we first remind the reader some basic features
of the quantum phase-space (Wigner) functions. We then generalize
the concept to the relativistic quarks and gluons in the proton.
With the preparation in Section II, the construction is
straightforward. However, the most general phase-space
distribution we define is not measurable at present, and hence we
proceed to make reductions by integrating out some dependent
variables. After integrating out the spatial coordinates, we
recover the transverse-momentum dependent parton
distributions\cite{Col03}. Upon integrating over the parton
transverse momentum, we have the reduced Wigner distributions
depending on 3-space coordinates and Feynman momentum $x$, which
are related to the GPDs by a simple Fourier transformation.
Therefore, the reduced quantum phase-space distributions are
physical observables.

\subsection{General Aspects of Wigner Distributions}

There is a vast literature on the quantum phase-space
distributions, and the Wigner distributions in particular. In
this subsection, we would like to summarize some of the salient
features.

Suppose we have a one-dimensional quantum mechanical system with
wave function $\psi(x)$, the Wigner distribution is defined as
\begin{equation}
W(x,p) = \int d \eta {\rm e}^{ip\eta} \psi^*(x-\eta/2)\psi(x+\eta/2)
\ ,
\label{QMWigner}
\end{equation}
where we have set $\hbar=1$. When integrating out the coordinate
$x$, one gets the momentum density $|\psi(p)|^2$, which is
positive definite. When integrating out $p$, the
positive-definite coordinate space density $|\psi(x)|^2$ follows.
For arbitrary $p$ and $x$, the Wigner distribution is not positive
definite and does not have a probability interpretation.
Nonetheless, for calculating the physical observables, one can
just take averages over the phase-space as if it is a classical
distribution
\begin{equation}
\langle \hat O(x,p)\rangle = \int dx dp \, W(x,p)O(x,p)
\end{equation}
where the operators are ordered according to the Weyl association
rule.
 For a single-particle system, the Wigner distribution
contains everything there is in the quantum wave function. For a
many-body system, the Wigner distribution can be used to
calculate the averages of all one-body operators. Sign changes in
the phase-space are a hint that it carries non-trivial quantum
phase information.

In the classical limit, the Wigner distribution is expected to
become classical phase-space distribution. For systems which are
statistical ensembles, the limit $\hbar \rightarrow 0$ is often
well-behaved. For example, for an ensemble of harmonic
oscillators at finite temperature, the Wigner distribution
becomes the classical Boltzmann distribution as $\hbar
\rightarrow 0$, see, e.g., \cite{DraHab98}. The Wigner
distribution for the $n$th excited state of the one-dimensional
harmonic oscillator of energy $E_n = \hbar \omega \left( n +
\frac{1}{2} \right)$ is \cite{Gro46}
\begin{equation}
W_n(p, x) = \frac{(-1)^n}{\pi\hbar} {\rm e}^{- 2 H /
(\hbar\omega)} L_n \left( \frac{4 H}{\hbar\omega} \right) \, ,
\end{equation}
where $H$ stands for the hamiltonian $H(p, x) = p^2/(2 m) + m
\omega^2 x^2/2$ and $L_n$ is the $n$th Laguerre polynomial. In the
quasi-classical limit---vanishing Planck constant and large
quantum numbers---the oscillator Wigner distribution turns into
the generalized distribution resided on the classical
trajectories $E_\infty = {\rm fixed}$ ,
\begin{equation}
\lim_{ \hbar \to 0, \, n \to \infty } W_n (p, x) \sim \delta \Big(
H (p, x) - E_\infty \Big) \, .
\end{equation}
Phase-space averaging with this kernel is equivalent to
calculating observables using classical equations of motion. This
can be easily understood from the semi-classical form of the wave
function
\begin{equation}
\psi (x) = C (x) {\rm e}^{i S (x)/\hbar} \, .
\end{equation}
Substituting this into Eq.\ (\ref{QMWigner}) and expanding $S$ to
the first order in $\hbar$, one gets the quasi-classical Wigner
distribution
\begin{equation}
W (p, x) = |C|^2 \delta \left( p - \frac{\partial S (x)}{\partial
x} \right) \, ,
\end{equation}
where the argument of the delta-function describes a family of
classical paths.

The quantum-mechanical Wigner distribution is measurable. The
actual measurement has been performed for a simplest quantum
system---the quantum state of a light mode (a pulse of laser
light of given frequency)---employing ideas of Vogel and Risken
\cite{VogRis89}. It was extracted via the method of homodyne
tomograhy \cite{SmiBacRayFar93} by measurement of a marginal
observable and subsequent reconstruction by the inverse Radon
transformation. Recently this Wigner distribution has been
measured directly by means of the photon counting techniques
based on a Mach-Zender interferometric scheme
\cite{BanRadWodKra99}.

Other versions of the phase-space distributions are possible.
They are all members of the so-called Cohen class \cite{Coh66},
with Husimi and Kirkwood distributions \cite{OthDis40} being its
well-known representatives. The Husimi distribution is a smeared
version of the Wigner distribution defined by projection of the
wave function on the coherent state (Gaussian wave packet)
\begin{eqnarray*}
H (\bar p, \bar x) = \int d p' d x' \, W (p', x') W_{\rm coh} (p'
- \bar p, x' - \bar x) \, ,
\end{eqnarray*}
which is real and positive-definite. On the other hand, the
Kirkwood function is complex. All these distributions are expected
to reduce to the same phase-space distribution in the $\hbar \to
0$ limit.

\subsection{Quantum Phase-Space Quark Distributions in the Proton}

In this subsection, we generalize the concept of phase-space
distributions to relativistic quarks and gluons in the proton. In
quantum field theory, the single-particle wave function must be
replaced by quantum fields, and hence it is natural to introduce
the {\it Wigner operator},
\begin{equation}
\hat{\cal W}_{\mit\Gamma} (\vec{r},k)
=
\int d^4\eta {\rm e}^{ik \cdot \eta}
{\mit \overline{\Psi}}(\vec{r}-\eta/2)
{\mit\Gamma} {\mit \Psi}(\vec{r}+\eta/2)
\ ,
\end{equation}
where $\vec{r}$ is the quark phase-space position and $k$ the
phase-space four-momentum conjugated to the spacetime separation
$\eta$. ${\mit\Gamma}$ is a Dirac matrix defining the types of quark
densities because the quarks are spin-1/2 relativistic particles.
Depending on the choice of ${\mit\Gamma}$, we can have vector, axial
vector, or tensor density.

Since QCD is a gauge theory, the two quark fields at different
spacetime points are not automatically gauge-invariant. One can
define a gauge-invariant quark field by adding a gauge link to
the spacetime infinity along a constant four-vector $n^\mu$,
\begin{equation}
{\mit\Psi} (\eta)= \exp\left(-ig\int^\infty_0 d \lambda \,
n \cdot A(\lambda n+\eta)\right) \psi(\eta) \ ,
\end{equation}
where we assume the non-singular gauges in which the gauge
potential vanish at the spacetime infinity
\cite{EfrRad80,LabSte85,ColSopSte88,Col02,BelJiYua02}. Clearly,
the Wigner operator depends on the choice of $n^\mu$. While
theoretically any $n^\mu$ is possible, experimentally $n^\mu$ is
constrained by the probes.

We have extended the Wigner distribution to including the time
variable. Therefore, beside the dependence on the 3-momentum,
there is also a dependence on the energy. For the bound states in
a simple system such as those in a simple harmonic oscillator, the
energy dependence is a $\delta$-function at the binding energies.
For many-body systems, however, the energy-dependence is more
complicated, as it reflects the distribution of the states after
one particle is removed from the system.

For non-relativistic systems for which the center-of-mass is
well-defined and fixed, one can define the phase-space
distributions by taking the expectation value of the above Wigner
operators in the $\vec{R}=0$ state. For the proton for which the
recoil effect cannot be neglected, the rest-frame state cannot be
uniquely defined, as discussed in Section II. Here we follow
Sachs, defining a rest-frame matrix element as that in the Breit
frame, averaging over all possible 3-momentum transfers.
Therefore, we construct the quantum phase-space quark
distribution in the proton as,
\begin{eqnarray}
{W}_{\mit\Gamma} (\vec{r}, k) &=& \frac{1}{2M_N}
\int \frac{d^3\vec{q}}{(2\pi)^3} \left\langle \vec{q}/2
 \left|\hat {\cal W}_{\mit\Gamma}(\vec{r}, k)
             \right|-\vec{q}/2\right\rangle  \\
                &=& \frac{1}{2M_N} \int \frac{d^3\vec{q}}{(2\pi)^3}
                    {\rm e}^{-i \vec{q}\cdot\vec{r}}
                   \left \langle \vec{q}/2\left|
          \hat {\cal W}_{\mit\Gamma} (0,k)\right|-\vec{q}/2\right\rangle
\ , \nonumber
\end{eqnarray}
where the plane-wave states are normalized relativistically. The
most general phase-space distribution depends on {\it seven}
independent variables.

The only way we know how to probe the single-particle
distributions is through high-energy processes, in which the
light-cone energy $k^-=(k^0-k^z)/\sqrt{2}$ is difficult to
measure, where the $z$-axis refers to the momentum direction of a
probe. Moreover, the leading observables in these processes are
associated with the ``good" components of the quark (gluon)
fields in the sense of light-cone quantization \cite{BroPauPin97},
which can be selected by ${\mit\Gamma} =\gamma^+$, $\gamma^+\gamma_5,$ or
$\sigma^{+\perp}$ where $\gamma^+=(\gamma^0+\gamma^z)/\sqrt{2}$.
The direction of the gauge link, $n^\mu$, is then determined by
the trajectories of high-energy partons traveling along the
light-cone $(1,0,0,-1)$ \cite{Col02,BelJiYua02}. Therefore, from
now on, we restrict ourselves to the reduced Wigner distributions
by integrating out $k^-$,
\begin{equation}
W_{\mit\Gamma}(\vec{r},\vec{k})
= \int \frac{dk^-}{(2\pi)^2} W_{\mit\Gamma}(\vec{r},k) \ ,
\label{wigner}
\end{equation}
with a light-cone gauge link is now implied. Unfortunately, there
is no known experiment at present capable of measuring this
6-dimensional distribution which may be called the {\it master} or
{\it mother} distribution.

Further phase-space reductions lead to measurable quantities.
Integrating out the transverse momentum of partons, we obtain a
4-dimensional quantum distribution
\begin{eqnarray}
\tilde f_{\mit\Gamma} (\vec{r},k^+) &=& \int \frac{d^2\vec{k}_\perp}
{(2\pi)^2}~ W_{\mit\Gamma}
 (\vec{r},\vec{k}) \nonumber \\
 & =& \frac{1}{2M_N} \int \frac{d^3\vec{q}}{(2\pi)^3}
 {\rm e}^{-i\vec{q}\cdot\vec{r}}  \int
 \frac{d\eta^-}{2\pi} {\rm e}^{i\eta^-k^+}
 \left\langle \vec{q}/2\left|
 \overline{\mit\Psi}(-\eta^-/2) {\mit\Gamma} {\mit\Psi}(\eta^-/2)
\right|-\vec{q}/2\right\rangle     \ .
\end{eqnarray}
The matrix element under the integrals is what defines the GPDs.
More precisely, if one replaces $k^+$ by Feynman variable $x p^+$
($p^+=E_q/ \sqrt{2}$, proton energy
$E_q=\sqrt{M^2+\vec{q}^2/4}$~) and $\eta^-$ by $\lambda/p^+$, the
reduced Wigner distribution becomes the Fourier transformation of
the GPD $F_{\mit\Gamma} (x, \xi, t)$
\begin{equation}
f_{\mit\Gamma} (\vec{r},x)
=
\frac{1}{2M_N}\int \frac{d^3\vec{q}}{(2\pi)^3}
{\rm e}^{-i\vec{q}\cdot\vec{r}}
F_{\mit\Gamma} (x, \xi, t) \ .
\label{dis}
\end{equation}
In the present context, the relation between kinematic variables
are $\xi=q^z/(2E_q)$ and $t=-\vec{q}^{~2}$. Taking $\Gamma =
\sqrt{2}\gamma^+$, the corresponding GPD has the expansion
\cite{Ji97}
\begin{eqnarray}
F_{\gamma^+}(x, \xi, t)
&=& \int \frac{d\lambda}{2\pi} {\rm e}^{i\lambda x}
               \left\langle \vec{q}/2\left|\overline{\psi}(-\lambda n/2)
                  {\cal L}\sqrt{2}\gamma^+
            \psi(\lambda n/2) \right|-\vec{q}/2\right\rangle \nonumber\\
     &=& H(x, \xi, t) \overline{U}(\vec{q}/2)\sqrt{2}\gamma^+U(-\vec{q}/2)
      + E(x, \xi, t) \overline{U}(\vec{q}/2)\frac{i\sigma^{+i}q_i}{ \sqrt{2}M}
       U(-\vec{q}/2) \ ,
\label{GPD}
\end{eqnarray}
where ${\cal L}$ is the shorthand for the light-cone gauge link.

The phase-space function $f_{\gamma^+}(\vec{r},x)$ can be used to
construct 3D images of the quarks for every selected Feynman
momentum $x$ in the rest frame of the proton. These images
provide the pictures of the proton seen through the Feynman
momentum (or ``color" or x) filters. They also may be regarded as
the result of a quantum phase-space tomography of the proton.
 We remind the reader again
that the Feynman momentum in the rest-frame sense is a special
combination of the off-shell energy and momentum along $z$,
namely $E + k^z$. Integrating over the $z$ coordinate, the GPDs
are set to $\xi\sim q^z=0$, and the resulting two-dimensional
density $f_{\gamma^+}(\vec{r}_\perp,x)$ is just the
impact-parameter-space distribution \cite{Bur00}. Further
integration over $\vec{r}_\perp$ recovers the usual Feynman parton
distribution.

The physical content of the above distribution is further revealed
by examining its spin structure. Working out the matrix element in
Eq. (\ref{GPD}),
\begin{eqnarray}
     \frac{1}{2M_N} F_{\gamma^+}(x, \xi, t) = \left[H(x,\xi,t) - \tau E(x,\xi,t)\right]
 + i [\vec{s}\times \vec{q}]^z
\frac{1}{2M_N}\left[H(x,\xi,t)+E(x,\xi,t)\right] ,
\end{eqnarray}
where $\tau = \vec{q}^2/4M^2_N$. The first term is independent of
the proton spin, and is considered as the phase-space charge
density
\begin{equation}
     \rho_+(\vec{r},x) =
   \int \frac{d^3\vec{q}}{(2\pi)^3} {\rm e}^{-i\vec{q}\cdot\vec{r}} [H(x,\xi,t)-
  \tau E(x,\xi,t)]   \ .
\end{equation}
The second term depends on the proton spin and can be regarded as
the third component of the phase-space vector current
\begin{eqnarray}
     j_{+}^z(\vec{r}, x) =  \int \frac{d^3\vec{q}}{(2\pi)^3}
     {\rm e}^{-i\vec{q}\cdot\vec{r}}
         i [\vec{s}\times \vec{q}]^z  \frac{1}{2M_N}\left[H(x,\xi,t)+E(x,\xi,t)\right] \ .
\end{eqnarray}
The $E$-term generates a convection current due to the orbital
angular momentum of massless quarks and vanishes when all quarks
are in the $s$-orbit. The physics in separating $f_\gamma^+$ into
$\rho_+$ and $j_+^z$ can be seen from the Dirac matrix $\gamma^+$
selected by the high-energy probes, which is a combination of
time and space components. Because the current distribution has
no spherical symmetry, the quark charge seen in the infinite
momentum frame, $\rho_++j_+^z$, is deformed in the impact
parameter space \cite{Bur02a}. This is the kinematic effect of
Lorentz boost.

Integrating the phase-space charge distribution
$\rho_+(\vec{r},x)$ over $x$, one recovers the
spherically-symmetric charge density in space. On the other hand,
if integrating over $x$ in $j_{+}^z(\vec{r},x)$, one obtains the
electric current density. In the latter case, if the integral is
weighted with $x$, one obtains the mechanical momentum density
\cite{Ji03}.

Finally, when integrating over $\vec{r}$ in the reduced Wigner
distributions in Eq. (\ref{wigner}), one obtains the
transverse-momentum dependent parton distributions.
\begin{equation}
q (x, \vec{k}_\perp) = \frac{M_N}{\sqrt{2} p^+}
\int \frac{d^3 \vec{r}}{(2 \pi)^2} W_+ (\vec{r}, \vec{k})
\ .
\end{equation}
There is a lot of interesting physics associated with these
distributions which has been discussed recently in the
literature. For instance, in a transversely polarized proton, the
quark momentum distribution has an azimuthal angular dependence
\cite{Siv90,BroHwaSch01,JiYua02}. The so-called Siver's function
can produce a novel single-spin asymmetry in deep-inelastic
scattering. We will not pursue this topic here, except
emphasizing that they have the same generating functions as the
GPDs.

\section{Three-Dimensional Images of the Quarks in the Proton}

Once the GPDs are extracted from experimental data or lattice QCD
calculations \cite{Mat00,GadJiJun02,Goc03,Hag03}, the phase-space
distributions can be obtained by straightforward Fourier
transformations. Without a first-hand knowledge on the GPDs at
present, we may be able to learn some general features of the
phase-space distributions form GPD models.

The GPDs have been parametrized directly to satisfy various
constraints \cite{Bel00,GoePolVan01,Bel01}, including 1) the first
moments reducing to the measured form factors, 2) the forward
limit reproducing the Feynman parton distributions, 3) the
$x$-moments satisfying the polynomiality condition \cite{Ji98},
and 4) the positivity conditions \cite{Pob}. In the first
subsection, we introduce a new parametrization without assuming
factorized dependence on the $t$ and other variables.

The GPDs were first calculated in a realistic nucleon model---the
MIT bag model \cite{JiMelSon97}. They have also been calculated in
the chiral-quark soliton model \cite{PolWei99,Pet98}. Recently,
there are calculations in the quark models as well
\cite{Bof02,Sco02}. In the second subsection, we will consider
the Wigner distributions in the bag model.

\subsection{A GPD Parametrization}

A generalized parton distributions depend on three variables,
$x$, $\xi$, and $t$. The simplest way to satisfy the
polynomiality condition is to relate it to a double distribution
\cite{Mue94,Rad99} and the $D$-term \cite{PolWei99}
\begin{equation}
\label{GPD-DD} H (x, \xi, t) = \int_{- 1}^{1} \frac{d y}{\xi} \,
{\mit\Xi} (y| x, \xi) F \left( y, \frac{x - y}{\xi}, t \right) +
\theta (\xi > |x|) D \left( \frac{x}{\xi}, t \right) \, ,
\end{equation}
The ``step"-function kernel in Eq.\ (\ref{GPD-DD}) has the form
\begin{eqnarray}
{\mit\Xi} (y| x, \xi) &=& \theta (x > \xi) \theta \left( \frac{x
+ \xi}{1 + \xi} \geq y \geq \frac{x - \xi}{1 - \xi} \right)
\nonumber\\
&+&\!\!\! \theta (- \xi > x) \theta \left( \frac{x + \xi}{1 - \xi}
\geq y \geq \frac{x - \xi}{1 + \xi} \right) + \theta (\xi
> |x|) \theta \left( \frac{x + \xi}{1 + \xi} \geq y \geq \frac{x -
\xi}{1 + \xi} \right) \, . \nonumber
\end{eqnarray}
The $q$-flavor double distribution $F_q = F^{\rm val}_q + F^{\rm
sea}_q$, including both valence and sea, can be related to the
non-forward quark distribution $f_q(y,t)$ through a profile
function $\pi(y,z,b)$,
\begin{eqnarray}
&& F^{\rm val}_q (y, z, t) = f^{\rm val}_q (y, t) \theta (y) \pi
(|y|, z; b_{\rm val})
\, , \\
&& F^{\rm sea}_q (y, z, t) = \left( \bar f_q (y, t) \theta (y) -
\bar f_q (- y, t) \theta (- y) \right) \pi (|y|, z; b_{\rm sea})
\, ,
\end{eqnarray}
where at $t= 0$ the function $f_q(y,t= 0)$ reduces to the
conventional parton distribution functions. The profile function
with a single parameter $b$ is assumed to be universal for
valence- and sea-quark species, and reads \cite{Rad99}
\begin{equation}
\label{DD-Ansatz} \pi (y, z; b) = \frac{{\mit\Gamma} \left( b +
3/2 \right)}{\sqrt{\pi} {\mit\Gamma} (b + 1)} \frac{\left[ (1 -
y)^2 - z^2 \right]^b}{(1 - y)^{2 b + 1}} \, .
\end{equation}

To proceed further, we design a non-factorized ansatz
\cite{GoePolVan01,BelMul02,Bur02a} for the functions $f_q (y, t)$
with intertwined $t$ and $y$ dependence. This is opposed to a
factorized form of GPDs with completely disentangled dependence
of the momentum transfer $t$ and scaling variables $(x, \xi)$. The
latter is currently accepted in almost all evaluations of physical
observables \cite{Bel00,Bel01,Fre03}. Due to a limited
kinematical coverage in the $t$-channel momentum transfer $t$ in
experiments, theoretical estimates confronted to data are
currently insensitive to this feature. Our model will be based on
the GRV leading order quark distributions \cite{GluReyVog98} with
discarded flavor asymmetry of the sea and it reads
\begin{eqnarray}
\label{NonfactorGPD} f^{\rm val}_{u} (y, t) &=& 1.239 y^{-
\alpha_v - \alpha'_v (1 - y)^{1/2} t} \left( 1 - 1.8 \sqrt{y} +
9.5 y \right) (1 - y)^{2.72}
\, , \\
f^{\rm val}_{d} (y, t) &=& 0.761 y^{- \alpha_v} \left( 2 y^{-
\alpha'_v (1 - y)^{1/2} t} - y^{- \beta'_v (1 - y) t} \right)
\left( 1 - 1.8 \sqrt{y} + 9.5 y \right) (1 - y)^{3.62}
\, , \nonumber\\
\bar f_{u} (y, t) &=& \bar f_{d} (y, t) = 0.76 y^{- \alpha_s -
\alpha'_s (1 - y)^{3/2} t} \left( 1 - 3.6 \sqrt{y} + 7.8 y
\right) (1 - y)^{9.1} \, . \nonumber
\end{eqnarray}
These models naturally reduce to the quark form factors with the
dipole parametrization of proton and neutron Sachs form factors.
The valence $d$-quark function has a more complicated structure
since the corresponding form factor $F_1^d$ has a node at $|t|
\approx 4 M^2/|2 \kappa_n + \kappa_p + 1|$: it is positive below
this value and is negative above it. The Regge intercepts and
slope parameters are taken as
\begin{eqnarray}
&&\alpha_v = 0.52 \, , \qquad \alpha'_v = 1.1 \, {\rm GeV}^{-2}
\, , \qquad \beta'_v = 1.0 \, {\rm GeV}^{-2}
\, , \\
&&\alpha_s = 0.85 \, , \qquad \alpha'_s = 0.3 \, {\rm GeV}^{-2}
\, . \nonumber
\end{eqnarray}
The valence quarks Regge parameters are numerically close to the
ones of $\rho$-reggeons, while the sea quarks being generated by
gluon radiation are analogous to the one of the pomeron. The form
factor asymptotics at large $t$ is governed by the large-$y$
behavior of $f (y, t)$. If the latter has the form $f (y, t) \sim
y^{- \alpha - \alpha' (1 - y)^p t} (1 - y)^N$ then the
corresponding form factor is $F (t \to \infty) \sim |t|^{- (N +
1)/(p + 1)}$. The perturbative QCD asymptotics for valence quarks
requires $p = 1$. We use however $p = 1/2$ for them since this
value fits better the form factor at small and moderate $t$. For
$p = 1$ one can get a decent behavior at moderate $t$ with
$\alpha'_u = 1.6 \, {\rm GeV}^2$. We use in our estimates $b_{\rm
val} = b_{\rm sea} = 1$. The $D$-term is parametrized as
\begin{equation}
D (z, t) = \left(1 - \frac{t}{m_{\rm D}^2} \right)^{- 3} (1 - z^2)
\left( d_0 \, C_1^{3/2} (z) + \cdots \right) \, ,
\end{equation}
with the mass scale $m_{\rm D}^2 = 0.6 \, {\rm GeV}^2$ and the
parameter $d_0$ computed within the $\chi$QSM
\cite{Pet98,GoePolVan01} and on the lattice \cite{Goc03,Hag03}
with the results
\begin{equation}
d_0^{\chi\rm QSM} = - 4.0 \frac{1}{N_f} \, , \qquad d_0^{\rm
latt} = d_0^u \approx d_0^d \approx - 0.5 \, ,
\end{equation}
respectively, where $N_f$ is the number of active flavors. In the
lattice case, the effect of disconnected diagrams was not
calculated, however they are known to produce a sizable negative
contribution \cite{GadJiJun02}. Once the latter are properly
taken into account the lattice result might approach the model
calculation. For our present estimate we chose an intermediate
value $d_0 = - 1.0$.

\begin{figure}[t]
\begin{center}
\mbox{
\begin{picture}(0,340)(250,0)
\put(0,190){\insertfig{5.3}{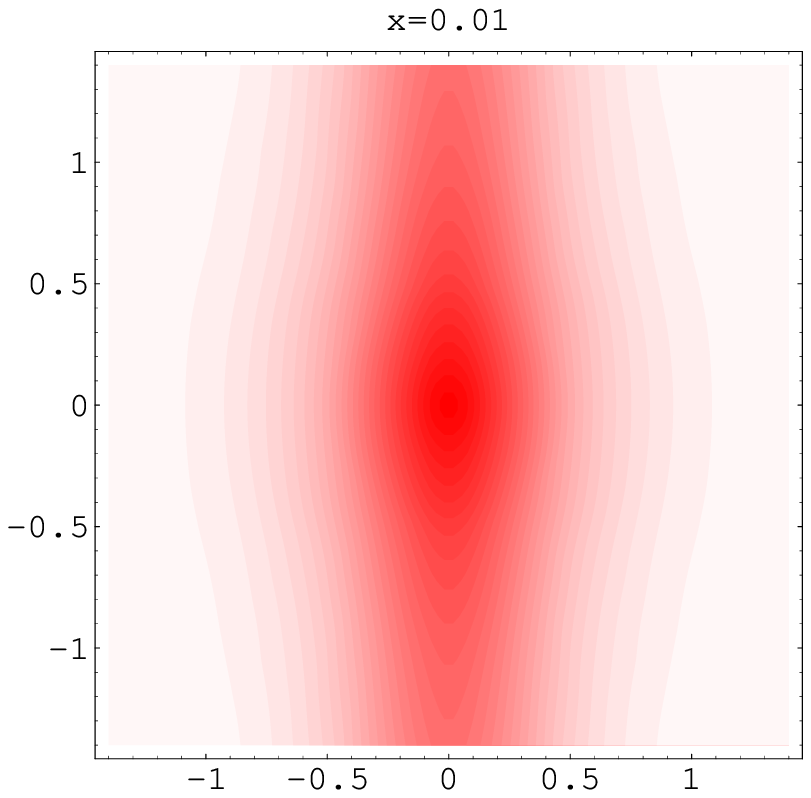}}
\put(170,190){\insertfig{5.3}{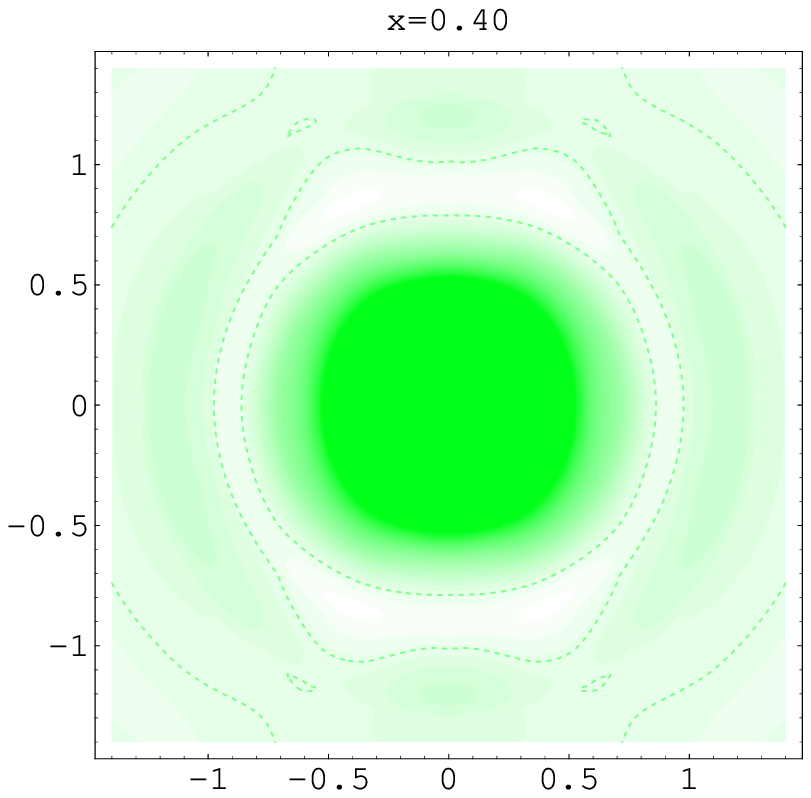}}
\put(340,190){\insertfig{5.3}{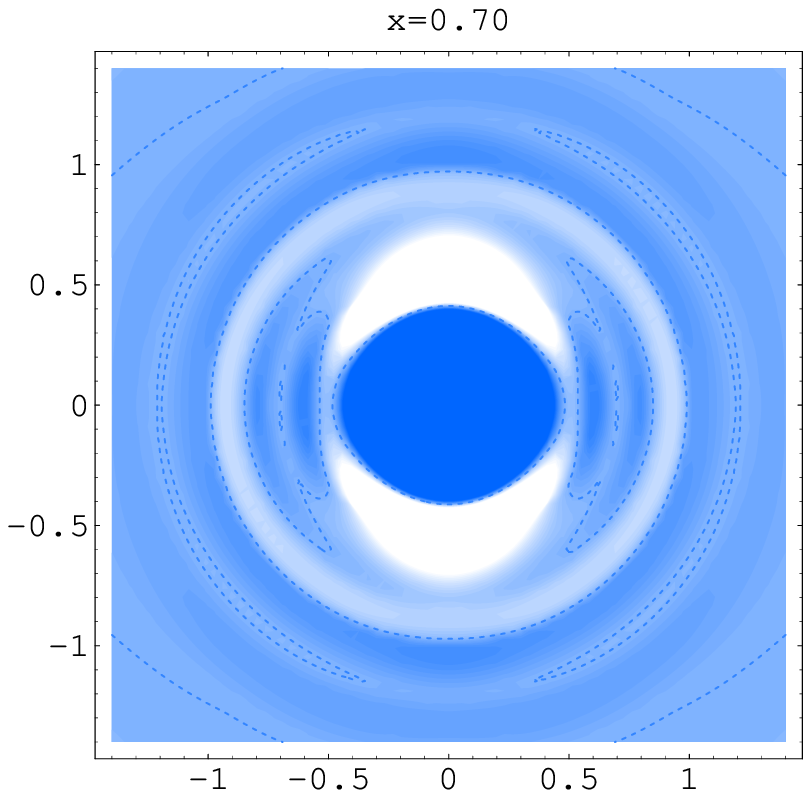}}
\put(0,0){\insertfig{5.3}{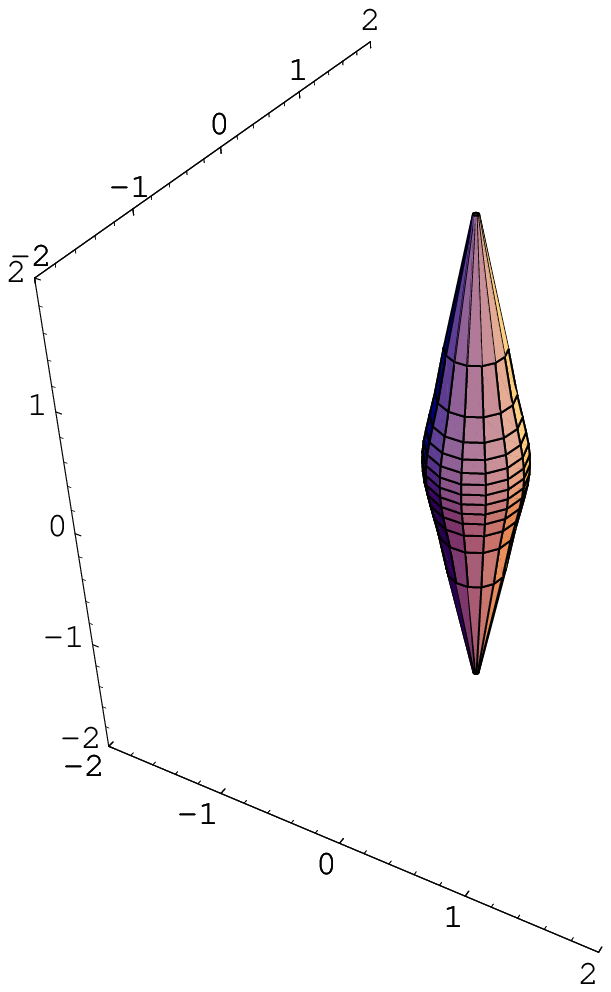}}
\put(170,0){\insertfig{5.3}{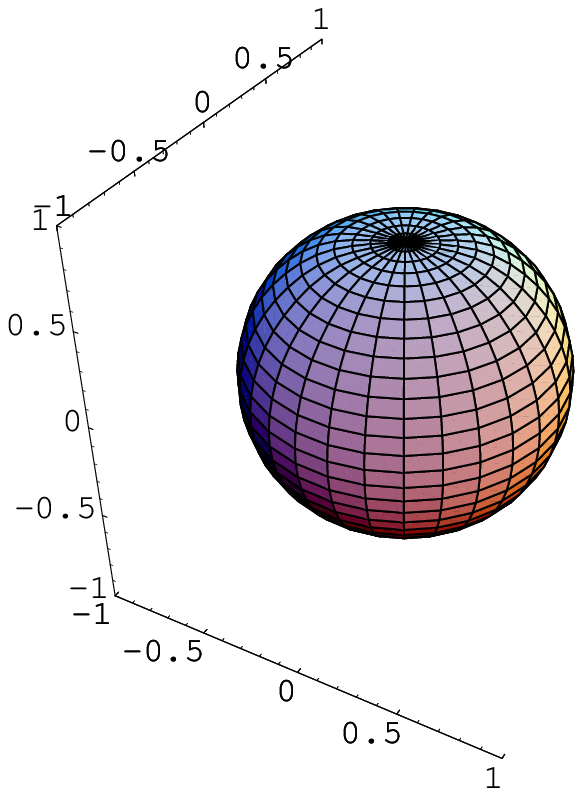}}
\put(340,0){\insertfig{5.3}{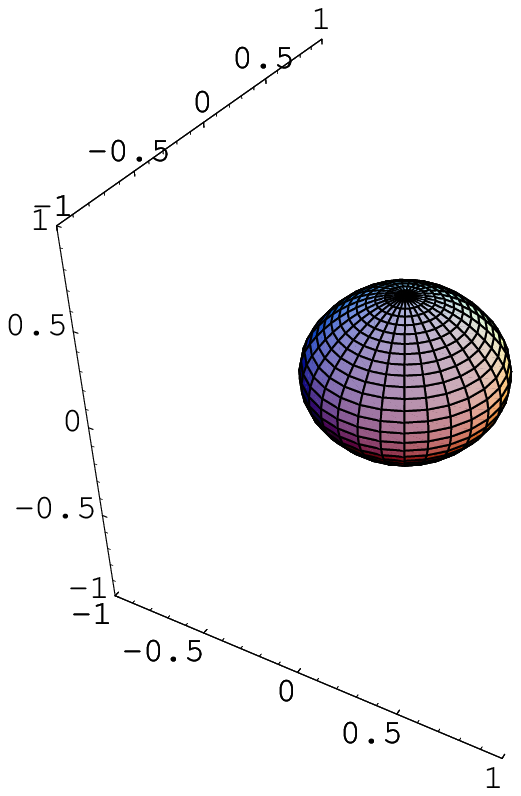}}
\end{picture}
}
\end{center}
\caption{\label{uQuarkNF} The $u$-quark phase-space charge
distribution at different values of the Feynman momentum for
non-factorizable ansatz of generalized parton distributions
(\protect\ref{NonfactorGPD}). The vertical and horizontal axis
corresponds to $z$ and $|\vec{r}_\perp|$, respectively, measured
in femtometers. The [dashed] contours separate regions of positive
[darker areas] and negative [lighter areas] densities. Below each
contour plot we presented the shape of three-dimensional
isodensity contours [$\rho = {\rm const}$].}
\end{figure}

According to the previous section, the phase-space charge
distribution $\rho_+(\vec{r},x)$ is just the Fourier
transformation of the above GPDs,
\begin{equation}
\rho_+^q(\vec{r},x)=\int
\frac{d^3\vec{q}}{(2\pi)^3} {\rm e}^{-i\vec{q}\cdot\vec{r}}
    H^q(x,\xi,t) \ ,
\end{equation}
where $\xi=q_z/2E_q$, $E_q=\sqrt{M^2+\vec{q}^2/4}$, and
$t=\vec{q}^2$. In the following, we consider the result of the
quark densities from the above parametrization.

In Fig.\ \ref{uQuarkNF} we show the up-quark charge distributions
calculated from $H_u (x, \xi, t)$ for various values of $x = \{
0.01, 0.4, 0.7 \}$. While the intensity of the plots indicates
the magnitude of the positive distribution, the lighter areas
below the ground-zero contours indicate negative values. The
plots show significant change in the distribution on the
longitudinal momentum fraction $x$. The image is rotationally
symmetric in the $\vec{r}_\perp$-plane. At small $x$, the
distribution extends far beyond the nominal nucleon size along
the $z$ direction. The physical explanation for this is that the
position space uncertainty of the quarks is large when $x$ is
small, and therefore the quarks are de-localized along the
longitudinal direction. This de-localization reflects a very
peculiar part of the nucleon wave function and shows long-range
correlations as verified in high-energy scattering. In a nucleus,
the parton distributions at small $x$ are strongly modified
because of the spatial overlap between the nucleons. On the other
hand, at larger $x$, the momentum along $z$ direction is of order
nucleon mass, the quarks are localized to within $1/M_N$. The
quantum mechanical nature of the distribution becomes distinct
because there are significant changes in the sign at different
spatial regions.

\begin{figure}[t]
\begin{center}
\mbox{
\begin{picture}(0,120)(250,0)
\put(0,0){\insertfig{4.3}{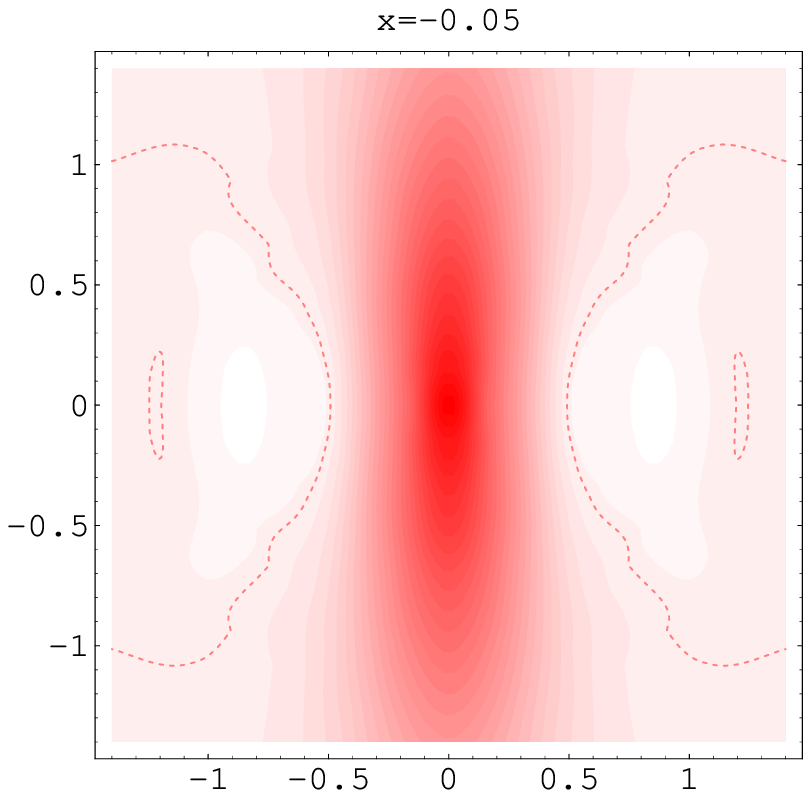}}
\put(125,0){\insertfig{4.3}{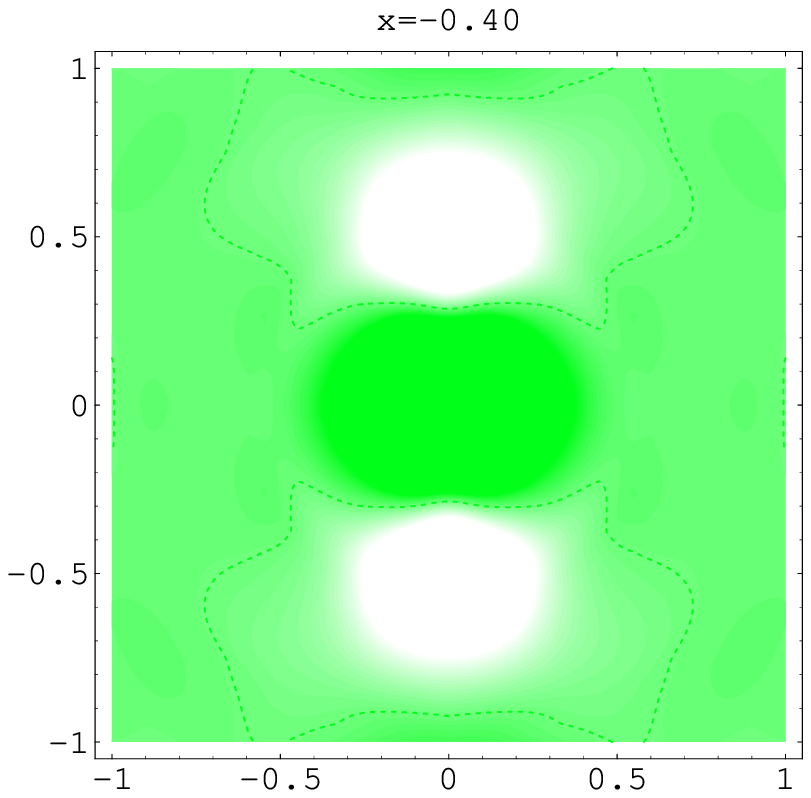}}
\put(250,0){\insertfig{4.3}{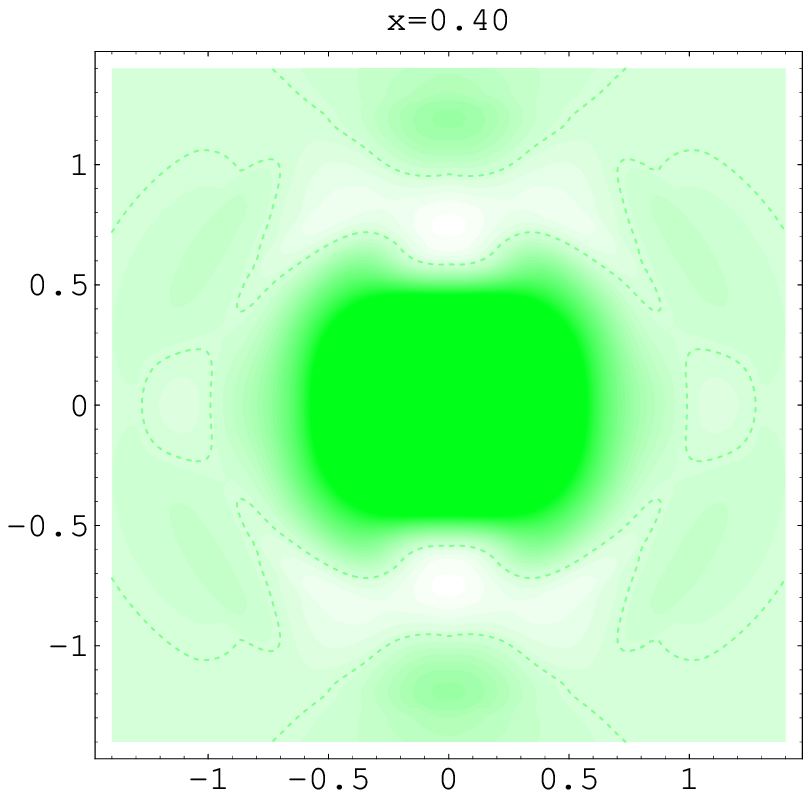}}
\put(375,0){\insertfig{4.3}{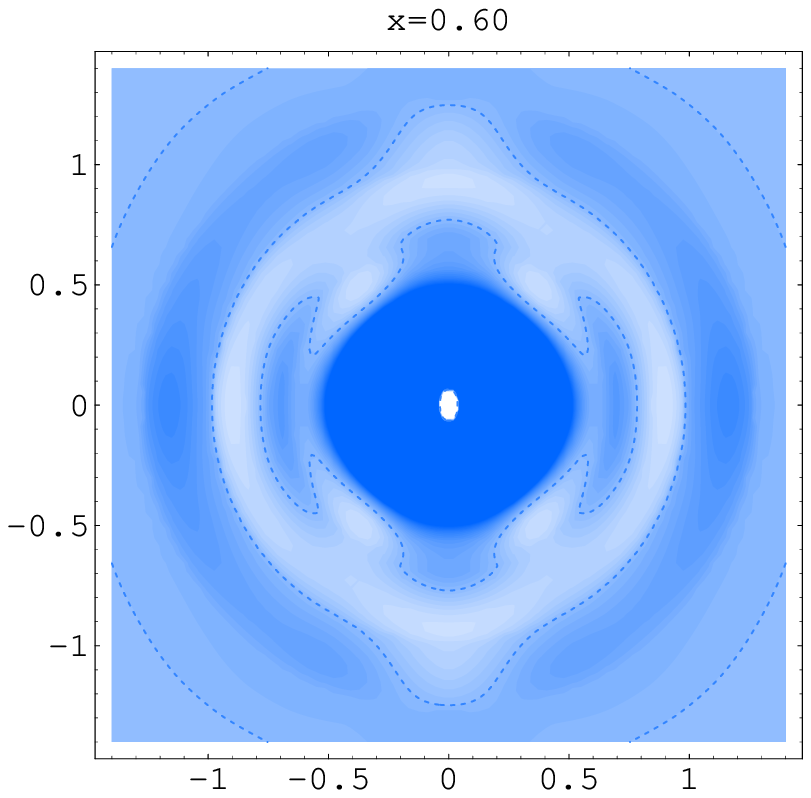}}
\end{picture}
}
\end{center}
\caption{\label{Others} The phase-space charge distribution for
the $u$-quark at negative Feynman momentum $x = -0.05$ and $x = -
0.4$ [two left panels] and $d$-quark for positive $x = 0.4$ and
$x = 0.6$ [two right panels].}
\end{figure}

It is also interesting to explore the distribution at negative
$x$. We show in Fig.\ \ref{Others} [two left panels] the Wigner
densities for the up-quark in the proton for $x$: $-0.05$ and
$-0.4$. These plots show significantly-different pattern than
those of the positive $x$. Finally, the two right panels we show
the density for the down-quark in the proton. The essential
features are quite similar to those of the up-quark densities.

\subsection{The MIT Bag Model}

The MIT bag model was invented more than a quarter of a century
ago \cite{Cho74}. The model was motivated by the color confinement
property of QCD. Massless quarks are confined to a cavity of
radius $R$, and move freely inside. The quark wave function is
ultra-relativistic and can be solved from the free Dirac equation
with spherical boundary conditions. The bag model has been used
to calculate many static properties of the nucleon and has had
many notable successes. The model can also be used to describe
the excitation spectrum of hadrons \cite{Cho74}. The
electromagnetic form factors \cite{BetGol83} and parton
distributions have also been calculated for the bag quarks
\cite{Jaf75}.

The bag model has been used to calculate the GPDs in Ref.
\cite{JiMelSon97}, where the boosted bag wave function has been
constructed using a simple prescription. In principle, one can
perform a Fourier transformation of the GPDs to calculate the
bag-model Wigner distribution.

However, we choose a simpler way to calculate the Wigner
distribution because the static bag has a fixed center. In fact,
we can calculate directly from the wave function of quarks in the
static nucleon just like in non-relativistic quantum mechanics.
The rational for this is that assuming the GPDs are known, one
can ``correct" the relativistic effects associated with the
boosted nucleon to obtain a Wigner distribution corresponding to
the static structure, just like applying the relativistic
corrections to extracting the static charge distributions. The
Wigner distributions calculated from a static bag correspond to
the ones with some relativistic corrections applied.

As we have discussed in the last section, we define the Wigner
distributions by the matrix elements of the Wigner operators
${\cal W}_+(\vec{r},k^+)$ in the hadron states. In general,
because of translational invariance, only the off-diagonal matrix
elements between nucleon states with finite momentum differences
provide the 3D $\vec{r}$ dependence. However, in the static models
such as the MIT bag, the quark wave functions are solved in the
rest frame of the nucleon which has no translational invariance
from start. With them, the Wigner distributions can be calculated
as the diagonal matrix elements of the Wigner operator for the
model nucleon fixed at the origin of the coordinates. For example,
\begin{equation}
\rho_+(\vec{r},x)=\frac{1}{2}\int\frac{d\lambda}{2\pi}
{\rm e}^{ix\lambda}
    \langle \vec{R}=0|\overline{\mit\Psi}(\vec{r}-(\lambda/2) n^-)\gamma^+
        {\mit\Psi}(\vec{r}+(\lambda/2) n^-)|\vec{R}=0\rangle \ ,
\label{bagd1}
\end{equation}
where $|\vec{R}=0\rangle$ represents the bag-model nucleon at
$\vec{R}=0$ and $x=k^+/p^+$ the light-cone momenta fraction of
the proton carried by the quark, $n$ a light-light vector with
$n^+=0, n^-=1/p^+, n_\perp=0$.

The quark field has the following expansion in the bag
\cite{Cho74}
\begin{equation}
\label{bw}
{\mit\Psi}_\alpha(\vec{r},t)=\sum\limits_{n>0,\kappa=\pm 1,m}
N(n\kappa) \{ b_\alpha(n\kappa m)\psi_{n\kappa
j=1/2m}(\vec{r},t)+ d_\alpha^\dagger(n\kappa m)\psi_{-n-\kappa
j=1/2m}(\vec{r},t)\} \ ,
\end{equation}
where $b_\alpha^\dagger$ and $d_\alpha^\dagger$ are the quark and
anti-quark creation operators in the bag, and $N(n\kappa)$ is a
normalization factor. The wave function are solved from the Dirac
equation with the bag boundary condition. For $j=1/2$ and
$\kappa=-1$, one has
\begin{equation}
\psi_{n,-1,\frac{1}{2}m}(\vec{r},t)=\frac{1}{\sqrt{4\pi}} \left(
\begin{array}{l}
i j_0(\frac{\omega_{n,-1}|\vec{r}|}{R_0})\chi_m\\
-\vec{\sigma}\cdot\hat{{r}}
j_1(\frac{\omega_{n,-1}|\vec{r}|}{R_0})\chi_m
\end{array}
\right)
{\rm e}^{-i\omega_{n,-1} t/R_0} \ .
\end{equation}
For the lowest mode, we have $n=1$, and $\omega_{1,-1}\approx
2.04$. In the above wave function, $\vec{\sigma}$ is the $2\times
2$ Pauli matrix, $\chi_m$ is the Pauli spinor, and $R_0$ is the
bag radius. $\hat{{r}}$ represents the unit vector in the
$\vec{r}$ direction, and $j_i$ are sphere Bessel functions.

\begin{figure}[t]
\begin{center}
\mbox{
\begin{picture}(0,120)(250,0)
\put(0,0){\insertfig{4.3}{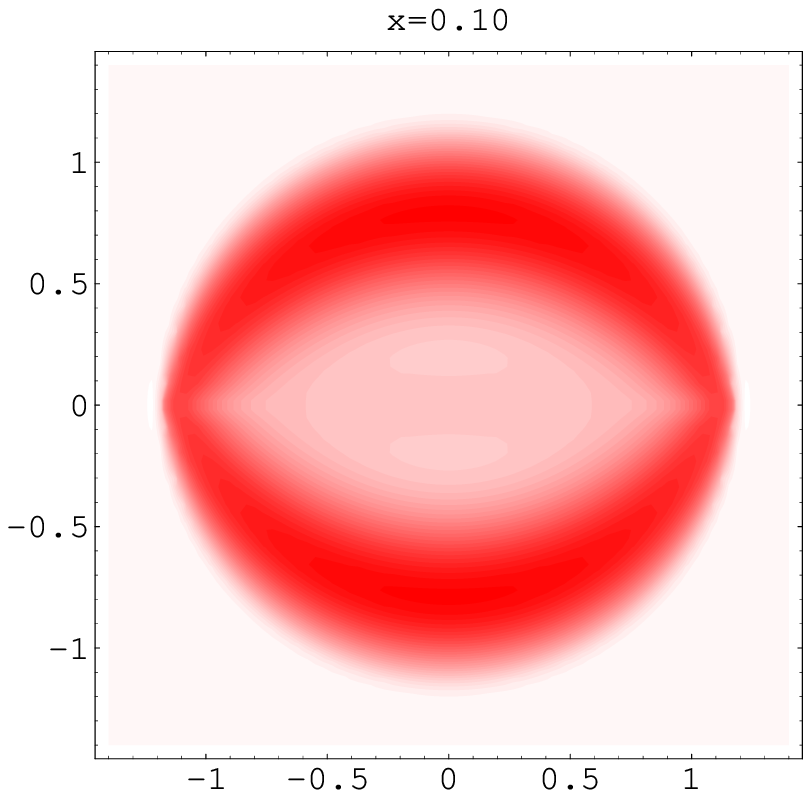}}
\put(125,0){\insertfig{4.3}{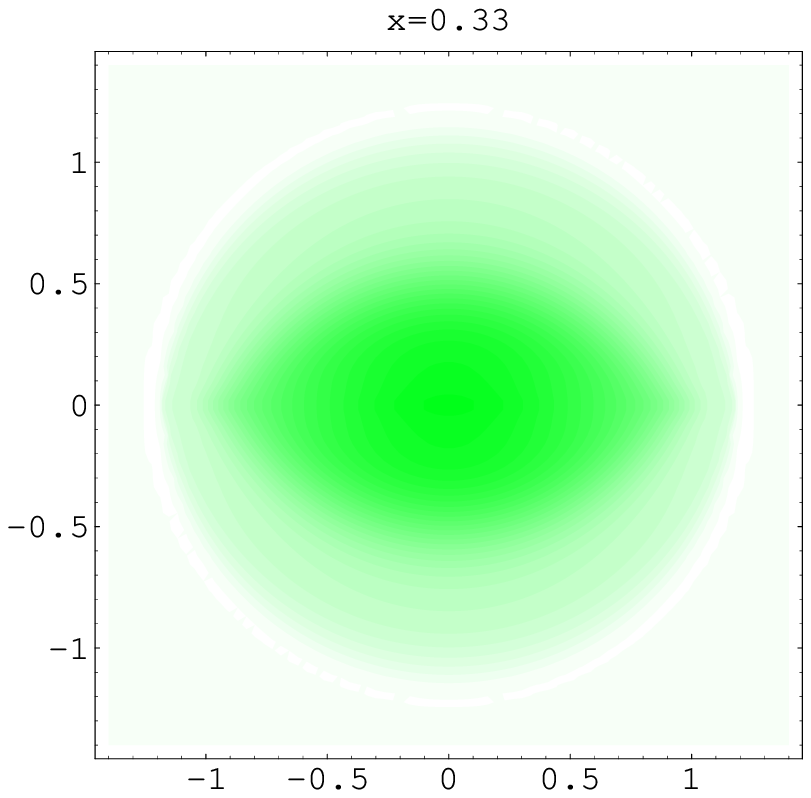}}
\put(250,0){\insertfig{4.3}{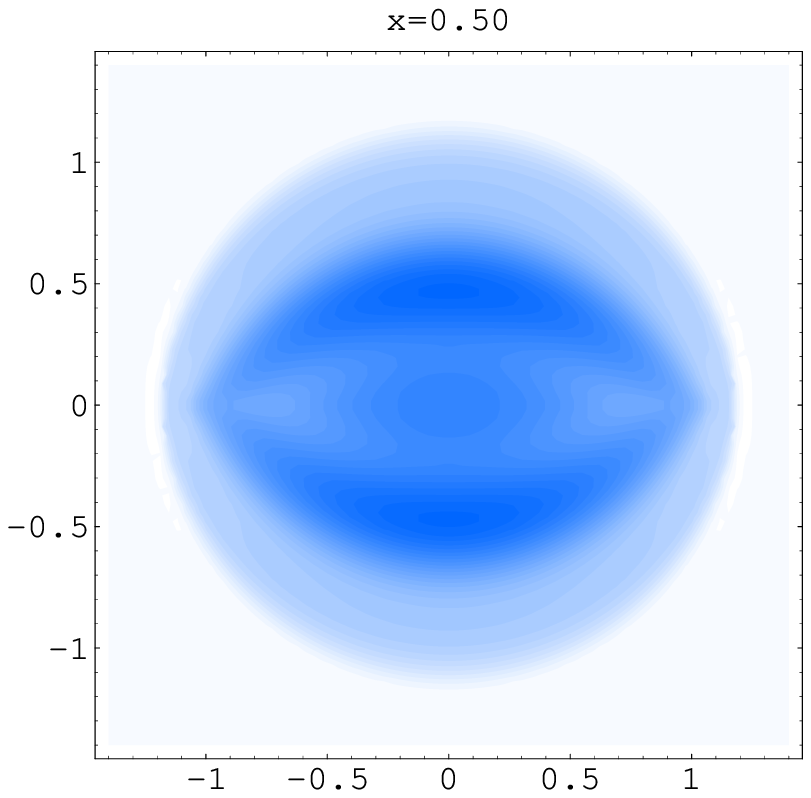}}
\put(375,0){\insertfig{4.3}{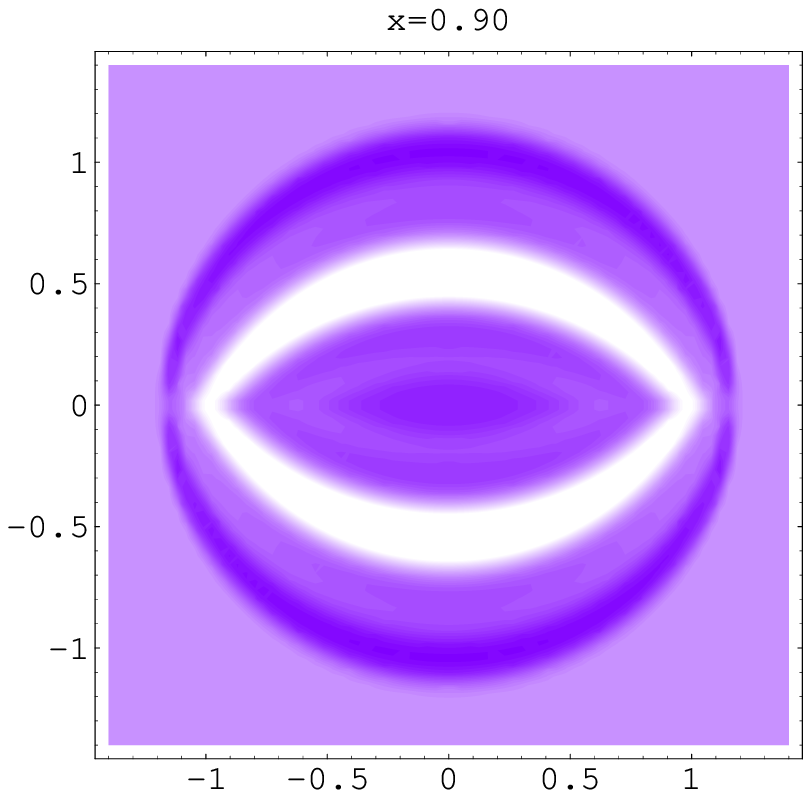}}
\end{picture}
}
\end{center}
\caption{\label{BagModel} The phase-space charge density $\rho_+
(\vec{r}, x)$ calculated in the bag model for values of Feynman
momentum $x = 0.1, 0.33, 0.5, 0.9$.}
\end{figure}

Substitute the above wave function into Eq.~(\ref{bagd1}), we
find the quark phase-space charge density,
\begin{eqnarray}
\rho_+^f(\vec{r},x)&=&C_f\frac{N^2}{4\pi}\int
\frac{d\lambda}{2\pi}
{\rm e}^{i\lambda(x-\frac{\omega}{MR_0})}
    \left[j_0(r_1)j_0(r_2)+j_1(r_1)j_1(r_2)\hat{r}_1\cdot\hat{r}_2
     \right. \nonumber\\
    &&\qquad\qquad\qquad\left. +
    i\left(j_0(r_1)j_1(r_2)\hat{r}_2^z-j_0(r_2)j_1(r_1)\hat{r}_1^z\right)\right] \ ,
\end{eqnarray}
where $C_f$ is a flavor factor with $C_u=2$ and $C_d=1$ for up and
down quarks, respectively. The position vectors $\vec{r}_1$ and
$\vec{r}_2$ are
\begin{equation}
\vec{r}_1=\vec{r}+\frac{\lambda}{2}\frac{1}{\sqrt{2} p^+} \hat e^z
\, ,
\qquad
\vec{r}_2=\vec{r}-\frac{\lambda}{2}\frac{1}{\sqrt{2} p^+}\hat e^z
\ .
\end{equation}
The above distribution satisfies the boundary constraint:
Integrating over $\vec{r}$ yields the quark distribution function,
while integrating over $x$ gives the charge density of the quarks
inside the nucleon.

With $\rho_+(\vec{r},x)$, one can visualize the quark charge
density as the function of $x$. In Fig. \ref{BagModel}, we have
shown a sequence of densities at $x=0.1, 0.33, 0.5$ and $0.9$. As
the parton density indicates, the charge density is peaked around
$x=1/3$ where the distribution is roughly spherical-symmetric.
This is consistent with the finding that the bag model GPDs have a
small $\xi$ dependence. For smaller and larger $x$, the charge
density can be negative. As $x$ increases, the distribution at the
center of the bag becomes smaller. As $x$ further increases, the
density there becomes negative. Similar phenomena happens as $x$
decreases. Because the bag boundary limits the distance of the
spatial correlation, the small-$x$ distribution does not grow
significantly as seen in experimental data.

\section{Summary and Conclusions}

In this paper, we have introduced the concept of the quantum
phase-space distributions for the quarks and gluons in the
nucleon. These distributions contain much more information than
conventional observables. In particular, various reductions of
the distribution lead to transverse-momentum dependent parton
distributions and generalized parton distributions.

 Any knowledge
on the GPDs can be immediately translated into the correlated
coordinate and momentum distributions of partons. In particular,
the GPDs can now be used to visualize the phase-space motion of
the quarks, and hence allow studying the contribution of the quark
orbital angular momentum to the spin of the nucleon.

In light of this, measurements of GPDs and/or direct lattice QCD
calculations of them will provide a fantastic window to the quark
and gluon dynamics in the proton.

We thank M. Burkardt and T. Cohen for many helpful discussions on
the subject of the paper, and M. Diehl for critical comments.
We thank the Department of Energy's Institute for Nuclear Theory at
the University of Washington for its hospitality during the program
``Generalized parton distributions and hard exclusive processes'' and
the Department of Energy for the partial support during the completion
of this paper. This work was supported by the U. S. Department of
Energy via grant DE-FG02-93ER-40762.

\end{document}